Dopant-defect interactions and their impact on local crystal stoichiometry studies in Mg-doped GaN via atom probe tomography

O. G. Licata[1], S. Broderick[1], E. Rocco[2], F. Shahedipour-Sandvik[2], and B. Mazumder[1]*

1. Department of Materials Design and Innovation, University at Buffalo, Buffalo, NY, USA.
2. College of Nanoscale Science and Engineering, SUNY Polytechnic Institute, Albany, NY, USA.

*Corresponding authors: baishakh@buffalo.edu*

**Abstract**

In this work, doping-defect interactions relevant to self-compensation in p-type GaN were investigated using atom probe tomography. The 3D visualization of ion distribution revealed the formation of spherical Mg-rich clusters and the segregation of Mg dopant towards dislocations in MOCVD-grown GaN:Mg. Impurities related to self-compensation, such as oxygen and hydrogen, were identified and detected adjacent to Mg-rich dislocations. Crystal stoichiometry around the defect regions was investigated to understand how the defects can serve as traps and influence dopant diffusion. Non-stoichiometric regions of N:Ga were found adjacent to Mg-rich dislocations and overlapping with some Mg-rich clusters, indicating potential traps. Variations in N:Ga were not proportional to the Mg content, suggesting that the microfeatures (clusters and dislocations) interact differently with local chemistry. Techniques for defining the quality of an APT experiment through invalidation of artifacts are also demonstrated. Mg-rich defects and variations in N:Ga were found to be independent of artifacts related to the evaporation field in APT.

**Introduction**

III-V systems, such as gallium nitride (GaN) have shown great promise in the field of optoelectronics and have achieved a wide spectrum range. Due to its large direct bandgap and strong interatomic bonds, GaN has been implemented in commercial blue light-emitting diodes (LEDs). A persistent limitation in the optimization of GaN-based optoelectronics is the relatively high p-type contact resistance. Magnesium (Mg) is the most effective p-type dopant; however, there is a significant drop-off in activation efficiency at high Mg doping concentrations, which are required to overcome the high ionization energy of Mg. Doping above ~2 x $10^{19}$ cm$^{-3}$ results in increased resistivity due to self-compensation of Mg acceptors.[1]

Self-compensation refers to the increasing concentration and decreased formation energy of point defects resulting from modulation of the bulk fermi level with dopant incorporation.[2] These point defects can play a role in the formation of extended defects, such as threading dislocations (TDs)[3]. The presence of TDs is known to degrade the efficiency, lifetime, and leakage current of III-N LEDs and laser diodes.[4] The distribution of Mg relative to defects has been shown to play a major role in optical performance.[5-7] The potential origin for self-compensation in GaN:Mg has been extensively investigated but is still not fully understood and is commonly attributed to one or a combination of the following mechanisms: (1) through the formation of electrically inactive Mg atoms within pyramidal inversion domains,[7-11] (2) through the formation of defect complexes containing nitrogen vacancies ($V_N$), such as the $Mg_{Ga}$-$V_N$ defect complex[12,13]. These defect complexes can be charged or neutral, which lowers the Fermi level and saturates free hole concentrations.[13,14] The introduction of defects suggests that there is degradation in the crystallinity and that this may also play a role in the compensation of holes. GaN layers grown on sapphire via metalorganic chemical vapor deposition (MOCVD) or metalorganic vapor-phase epitaxy (MOVPE) have carbon (C), oxygen (O), and hydrogen (H) impurities originating from metalorganic precursors and carrier gases.[15] C impurities have been shown to contribute to carrier compensation in GaN:Mg;[16-18] however, it is more energetically favorable for O to incorporate in N-polar GaN, while C is more favorable in Ga-polar GaN.[19,20] O impurities have been identified predominately along the cores of screw and mixed dislocations.[21,22] H reduces doping efficiency in as-grown samples and forms Mg-H complexes.[13,23,24] Much of the H introduced in MOCVD is removed following post-growth annealing; however, not all H is removed.[24-26] Mg-H complexes that remain after annealing passivate Mg and act as compensating centers.[24,27]

Many of our understandings on self-compensation mechanisms in GaN:Mg are based on computational models.[13,28-31] Experimental verification is often based on measurements taken at the sample scale, such as bulk or averaged concentration,[24,32] dislocation density,[33,34] or hole mobility.[1] There are also limitations in standard experimental techniques, such as the limited lateral resolution of SIMS and lack of depth sensitivity in TEM. There is a need for experimental verification at the scale of the defects themselves in order to truly understand the local interactions of defects, dopants, and impurities within GaN:Mg. Atom probe tomography (APT) easily lends itself to visualizing the three-dimensional (3D) distribution of Mg relative to defect structures, given its near-atomic spatial resolution and sensitivity to light elements at low concentration (APT limit is $\geq 10^{17}$ cm$^{-3}$).[35,36] In our previous work, APT revealed that Mg cluster density (clusters per ccm) and radius both decreased within the sidewalls of hexagonal hillock structures in N-polar GaN.[37] In the proposed work, APT is utilized to further investigate the relationship between Mg distribution and defects by including the influence of local crystal stoichiometry and impurity distribution.

**Methods**

The GaN:Mg structure was prepared via MOCVD; details on the growth and preparation were previously reported by Rocco et al.[37] and Marini et al.[38] APT samples were prepared in an FEI DualBeam 875 focused ion beam (FIB) through the standard wedge lift-out technique, then annularly milled to sharp needle-like tips with an end radius less than 100 nm.[39] Samples were chosen from site-specific regions relative to the hillock sidewall. APT acquisitions were carried out in a CAMECA, Inc. Local Electrode Atom Probe (LEAP®) 5000XR, equipped with a reflectron lens and ultraviolet ($\lambda = 355$ nm) laser pulsing capabilities. The APT analysis conditions consisted of 30 K base temperature, 4-5 pJ pulse energy, 0.5-0.8% detection rate, and 250 kHz pulse frequency. All APT analyses revealed segregation of Mg into clusters and towards dislocations. The GaN:Mg sample investigated in this work was a representative sample, taken from the start of the hillock sidewall. This particular sample was ideal given the presence of dense and continuous dislocations adjacent to but not entirely coincident with the sample tip's edge (where erroneous data is more likely to appear).

APT was employed to reveal the 3D distribution of dopant (Mg) and impurities (C, O, and H) relative to the bulk GaN and dopant-defect interactions. Principal component analysis was employed to define which physical parameters associated with field evaporation contributed to the variability. Additional measures relative to the field evaporation and elemental homogeneity, including the charge state ratio (Ga$^{++}$/Ga$^+$) and density of ions, were taken to reveal potential artifacts. Loss of nitrogen in APT is known to result in inaccurate stoichiometry in III-N compound materials.[40,41] For this reason, the standard practice is to report the III-site composition, excluding N.[35,42-44] Here, we report the ratio of N to Ga following normalization of the mean value to 1.

**Results & Discussion**

Atom probe tomography (APT) relies on time-of-flight mass spectrometry and the resulting mass spectrum is the foundation for all subsequent analyses. The full APT mass spectrum for GaN:Mg is shown in Figure 1a. The peaks for Mg$^{2+}$ are at 12, 12.5, and 13 Da, which can overlap with C$^+$ at 12 Da. Decomposition of the peaks based on the known isotopic abundancies for Mg did not reveal any C above the detectable limit ($10^{17}$ cm$^{-3}$). H and O species were identified in the mass spectrum as impurities and are included in Figures

1b-d. Typically, H species are difficult to accurately quantify in APT due to the presence of residual gases in the analysis chamber.[45] In Figure 1b, the hydride peaks at 1, 2, and 3 Da were included to qualitatively understand the known relationship between Mg and H, namely, Mg-H complex formation.[18,24,33] The elemental H peaks may contain a combination of real impurity events and residual gas. For this reason, it is common to exclude H from the reported composition. Exceptions to this practice include APT investigations where the material system demonstrated H-rich precipitates.[46,47] Here, H is included to provide a qualitative representation of potential inhomogeneity relative to Mg and the N:Ga crystal stoichiometry. Molecular hydride species were excluded from the impurity distribution to focus solely on the potential inhomogeneity of the H ion without influence from its interaction with other evaporated species. The three-dimensional atom map acquired through APT, including Ga atoms (blue) and Mg clusters (magenta), is shown in Figure 1e. Magenta isosurfaces were generated around Mg-rich clusters in the CAMECA, Inc. Integrated Visualization Analysis Software (IVAS). Isosurfaces have a designated threshold concentration value, referred to as the 'isovalue'. Due to the variation in Mg concentration between clusters and dislocations, Mg clusters are shown with an isovalue of 0.9 at.% Mg and dislocations with 1.4 at.% Mg. Two Mg-rich dislocations were identified and are shown in Figure 1e. Threading dislocations often form along the c-axis in GaN grown on sapphire due to the lattice mismatch.[48,49] Since Mg is a larger ion than Ga, it is expected to segregate to regions of increased tensile strain, including edge and mixed type threading dislocations.[4]

To ensure robust analysis, measurements were taken across multiple sample regions. For optimal assessment, any potential biases in the data, whether resulting from the material, the sample evolution or the evaporation physics were considered. For example, the presence of Mg-rich micro-features (clusters and defects) could potentially result in inhomogeneous field evaporation, thereby impacting the accuracy of the structure-chemistry analysis. In order to assess the suitability of the multiple regions, a principal component analysis (PCA) was performed on relevant descriptors. These descriptors encompass potential effects associated with the instrument (DC voltage; VDC), the evaporation physics (number of pulses between recorded events; $\Delta P$), the chemistry (time-of-flight; TOF), the structure (reduction of the position coordinates, X and Y, to the root sum squared; RSS), or a combination of effects such as structure and chemistry (like-nearest neighbor distance; Like-NN). Beyond the variables associated with evaporation physics extracted from the IVAS-derived EPOS file, two additional variables: RSS and Like-NN, were calculated and defined. Like-NN was computed by taking the average of the k-nearest neighbor distance, with k equal to 6. Here, Like-NN served as a measure of homogeneity and a means of correlating other variables with the segregation of species to clusters and dislocations.

PCA defines the correlation between descriptors, across a large number of incidents. Thus, it serves to identify the factors most impacting the results, with these impacts difficult to otherwise identify due to the massive data size resulting from APT experiments. The region of interest captured a volume consisting of approximately half Mg dislocation and half GaN:Mg matrix. Table 1 shows the resulting contribution to variability from the principal components (PCs). It is found that ΔP is the dominant change in the descriptor space, with greater than 70% of the variability resulting from that variable and, more critically given our classifications, the evaporation physics. Thus, this raises the potential issue that different regions may not be directly comparable. However, as shown in the Supplementary Section (and particularly Figure S1), the average value of ΔP along the sample tip does not show any variation resulting from the presence of Mg segregation. Additionally, 2D projections of the ΔP distribution appear random and homogeneous. Thus, the variability is not due to a change in physics but rather is reflective of noise. The next highest variable impact was for Like-NN, but that is attributed to Mg segregation. The other features in the analysis accounted for little variability. From this, we can conclude that different regions of the sample are indeed comparable because the variation falls within the range of noise and the evaporation-related phenomena do not influence the observed chemical inhomogeneity.

The presence of potential artifacts was thoroughly investigated by quantifying the homogeneity of the field evaporation within voxelized cross-sections. This was carried out by measuring the density of ions and relative field evaporation (charge-state-ratio) in a cross-section without Mg segregation compared to a cross-section with Mg clusters and dislocations. The details of this portion of the investigation are shown in Figure S2 of the supplementary. The density of ions was mostly random and did not correlate with the known Mg-rich features. The relative field evaporation was overall homogeneous for both regions. The variables investigated through PCA and voxelization of cross-sections indicated that the observed Mg-rich defects and variations in N:Ga did not result from artifacts. After verifying that the Mg-rich zones were independent of evaporation-related artifacts, all defect interaction studies could be reported confidently.

The impact of N, O, and H on the neighborhood stoichiometry of Mg-rich dislocations was investigated. A region of interest, measuring 30 x 12 x 80 nm$^3$, was chosen to encompass the entire dislocation (dislocation indicated with an arrow in Fig. 1e). 2D concentration profiles were generated for the three planes to show the 3D distribution. Figure 2a shows the resulting distribution of Mg in at.%. The edges of the Mg segregation are outlined in magenta and overlaid with the other species in Figures 2b-d. Figure 2b shows the normalized ratio of N to Ga, where the mean value of N:Ga was normalized to 1. Figure 2b demonstrates that high N:Ga occurs mostly to the left of the Mg-rich dislocation, with some overlap, and lower N:Ga to the right. Figure 2c shows the O:Ga ratio relative to the Mg-rich dislocation and reveals a few high O:Ga zones to the left of the Mg-rich zone. This finding agrees with the 'oxygen-decorated' dislocations described by Michalowski et al.[21] Figure 2d demonstrates that the distribution of elemental H:Ga is immediately to the left of the Mg-rich dislocation and less so at the edge of the sample or within the bulk. The proximity of the H-profile to the Mg-dislocation may indicate the presence of stable Mg-H complexes that are electrically inactive and remain after annealing.[24] The dislocations serve as a sink for Mg dopant ions; additionally, the adjacent crystal structure is disrupted, leading to trapped impurities and variations in N:Ga stoichiometry.

The composition adjacent to the Mg-rich dislocations was analyzed through 2D concentration profiles to visualize the inhomogeneity in-plane (XY plane). A lower region along the z-axis was chosen to capture the Mg-rich dislocations where they do not coincide with the edge of the sample. The atomic percent of Mg within the 2 nm thick volume is shown in Figure 3a. The corresponding N:Ga composition is displayed in Figure 3b. From the 2D concentration profiles, higher N:Ga values appear close to Mg and continue toward the sample edges. A 1D line profile was taken through both dislocations to investigate their relative concentrations. As shown in Figure 3c, the peak concentration of N:Ga is just beyond the edge of the Mg-rich zone. Figure 3d and Figure 3e show the line profiles of H:Ga and O:Ga, respectively, along with the Mg concentration. Both H:Ga and O:Ga show increased values near the Mg dislocation; however, there is also visible fluctuation in concentration due to their statistically low counts. It is important to note that, between the 25 to 55 nm region of the 1D line profile, the fluctuation of H:Ga and O:Ga is significantly reduced, and the impurity concentration is very low. The O:Ga adjacent to Mg-rich dislocation agrees with previous findings and further confirms the local segregation of Mg that could not be verified by SIMS.[21] The observed H-rich zones adjacent to the Mg-rich dislocation suggest the formation of Mg-H complexes that remained after annealing.[24] Diffusion along dislocations is much faster than within the bulk of crystal lattices due to the 'pipe diffusion' mechanism.[50,51] The disordered core of the dislocation is thought to lower the activation energy for diffusion.[50,52] The increased tensile strain and disorder of the dislocation core account for the segregation of Mg; however, they do not explain the segregation of impurities adjacent to the Mg-rich zones. This behavior suggests diffusion paths relative to the dislocations and adjacent degradation within the crystal structure. In the dislocation-mediated pipe diffusion process, impurities diffuse both through and out of pipes along the length of the defect.[53] This may explain the presence of impurities observed both along and adjacent to the dislocations. It has also been reported that pipe diffusion occurs as the result of strong strain fields not only along the 1D dislocation but in the strained lattice surrounding the dislocation.[54] Previously, Liliential-Weber et al. attributed excess impurities to the presence of a Ga-rich columns along screw dislocations.[55] Here, we do not observe an increase in Ga, relative to the bulk concentration.

A major challenge in defining the cause of the adjacent segregation is pinpointing when it occurred. In these samples, both the formation of dislocations and Mg doping occur during the growth process, i.e. the Mg is not introduced post growth. It is plausible that point defects form beyond the core of the dislocation and perhaps serve as traps for impurities. One of the potential mechanisms for self-compensation relies on the formation of defect complexes containing nitrogen defects.[13] The diffusion of N vacancies is predicted to be faster in directions perpendicular to the c-axis.[12] Here, the dislocations themselves are likely along the direction of the c-axis,[48] which shares an acute angle with the APT sample's z-axis due to the semi-polar nature of the hillock sidewall. Significant diffusion of $V_N$ is unlikely in this case; however, due to the low annealing temperature used (775°C).[56]

To define the role of Mg-rich clusters in trapping impurities compared to the dislocations, the detailed chemical analysis around those regions was performed in a similar manner. As shown in Figure 4a (i) and Figure 4b (i), two regions containing both Mg-rich dislocations and at least one Mg cluster were chosen for the analysis. Figure 4a (ii) and Figure 4b (ii) indicate the N:Ga relative to the Mg-rich zones. From these 2D concentration profiles and the 1D line profiles shown in Figure 4a (iii) and Figure 4b (iii), it is clear that the Mg cluster in Figure 4a is rich in N:Ga while the Mg cluster in Figure 4b has average N:Ga concentration. This suggests that Mg-rich dislocations and Mg-rich clusters interact differently with local chemistry. Increased N:Ga values are always found adjacent to the dislocations, but not all clusters act as trapping sites for N diffusion. Further inspection of Figure 4a (iii) and Figure 4b (iii) shows that the N:Ga ratio is not proportional to the atomic percent of Mg, i.e., the clusters captured in both regions are approximately 2 at.% Mg but they show drastically different ratios of N:Ga. The two Mg-rich dislocations demonstrate very different Mg concentrations and yet have similar N:Ga ratios. These findings confirm that the Mg content alone is not influencing the variation in N:Ga and rather suggests that the clusters and dislocations impact their local crystal structure differently. The variation in local stoichiometry may be an indicator for the ease of diffusion relative to specific defect zones as each ion species has a unique rate of diffusion.

**Conclusion**

In this work, dopant-defect interactions associated with self-compensation in p-type GaN were investigated at the atomic level. APT allowed for 3D-visualization of dopant (Mg) and impurity (O and H) distributions related to defects in the crystal stoichiometry of MOCVD-grown GaN:Mg. Mg-rich dislocations demonstrated a consistent trend of increased N:Ga adjacent to Mg. N:Ga relative to clusters varied, with some clusters showing increased N:Ga adjacent to Mg while others demonstrated direct overlap of high N:Ga with Mg. Variations in N:Ga were not proportional to the Mg content, suggesting that the micro-features (clusters and dislocations) interact differently with local chemistry. C impurities were not observed above the detectable limit of APT ($10^{17}$ cm$^{-3}$). O impurities were found adjacent to Mg-rich dislocations, which is consistent with previous findings via SIMS.[21] H-rich zones along Mg-rich dislocations suggest the presence of Mg-H complexes that remained after the post-growth annealing process. A thorough analysis of potential artifacts was performed to confirm that the Mg-rich defects and variations in N:Ga were independent of artifacts related to the evaporation field in APT. These findings reveal the influence of crystal stoichiometry defects on dopant and impurity distribution and how APT can accurately locate these defects through compositional analysis. Further investigations, including correlations with resistivity, will aid in the effort to understand hole compensation at the atomic level.

**Supplementary materials**
See supplementary materials for the spatial investigation of ΔP relative to Mg-rich micro-features and the homogeneity of the field evaporation (density of ions and charge-state-ratio).

**Availability of Data**
The data that supports the findings of this study are available within the article.

**Tables:**

Table. I. Principal component contribution to variability and corresponding input variable. The variables included a simplification of the position coordinate (RSS), time-of-flight (TOF), DC voltage (VDC), number of pulses between recorded events (ΔP), and the average distance between neighboring ions of the same species (Like-NN).

| Component | Contribution to variability (%) | Feature |
|---|---|---|
| PC1 | 71.35 | ΔP |
| PC2 | 15.75 | Like-NN |
| PC3 | 11.67 | RSS |
| PC4 | 1.19 | TOF |
| PC5 | 0.04 | $V_{DC}$ |

**Figure captions:**

Figure 1: (a) The labeled mass spectrum for GaN:Mg with insets magnifying major elements and impurities: (b) hydrides, (c) $Mg^{++}$, $N^+$, and $O^+$, and (d) $Mg^+$, $N_2^+$, and $O_2^+$. (e) The 3-D atom map including Mg-rich isosurfaces in dark pink and Ga atoms in blue. Ga concentration is reduced to 3.0% to aid in visualizing both Ga and Mg. A dislocation and cluster are labeled and indicated by arrows. The two dislocations have isosurface concentrations of 1.4 at.% Mg. All clusters have an isosurface concentration of 0.9 at.% Mg.

Figure 2: 2D composition profiles from an Mg-rich dislocation in the x, y, and z planes. Dashed lines show the location of Mg relative to the composition of (a) Mg, (b) N:Ga, (c) O:Ga, and (d) H:Ga.

Figure 3: 2D concentration profile of (a) Mg atomic percent and (b) normalized N:Ga in the xy plane. Two Mg-rich dislocations are indicated. The volume within the dashed box was used for the 1D line profiles in (c-e). 1D line profiles show the Mg concentration relative to (c) N, (d) O, and (e) H.

Figure 4: Two 2 nm thick slices in z were taken at (a) z = 255 nm and (b) z = 274. 2D concentration profile of (i) Mg atomic percent and (ii) normalized N:Ga in the xy plane. An Mg-rich cluster and dislocation are indicated. The volume within the dashed box was used for the 1D line profile in (iii).

**Figures:**

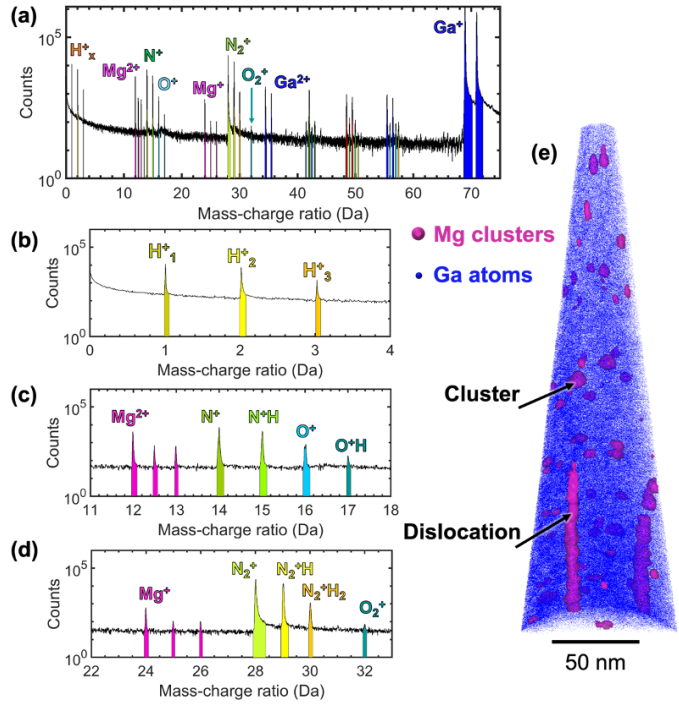

Figure 1

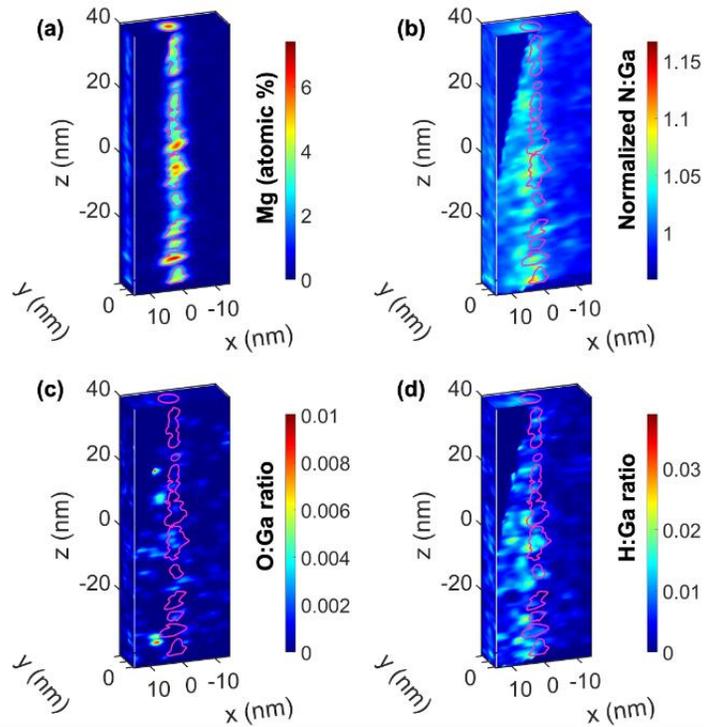

Figure 2

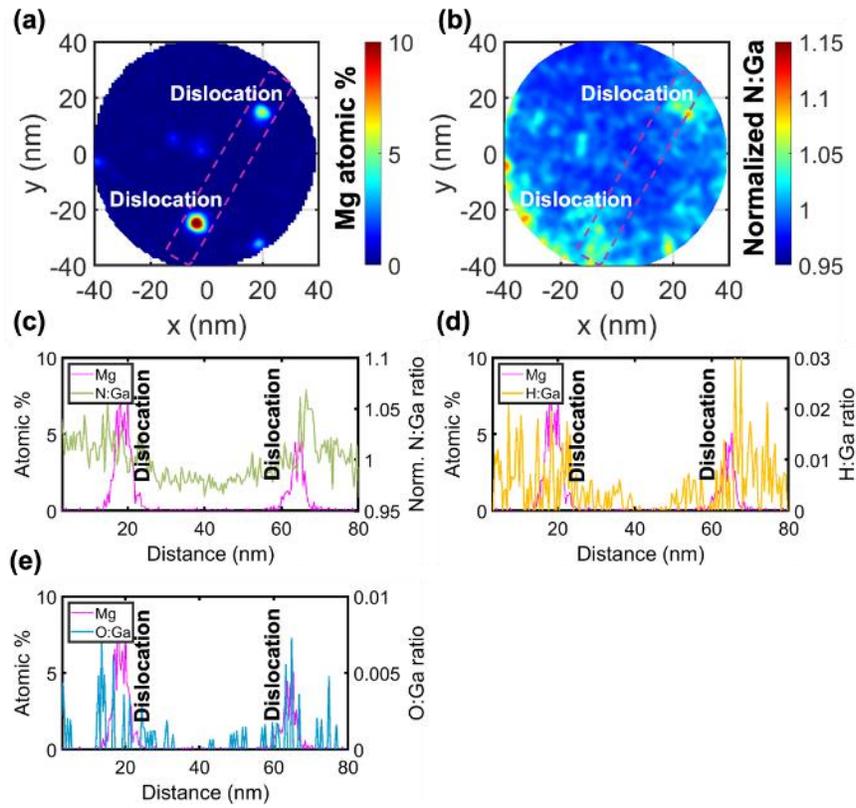

Figure 3

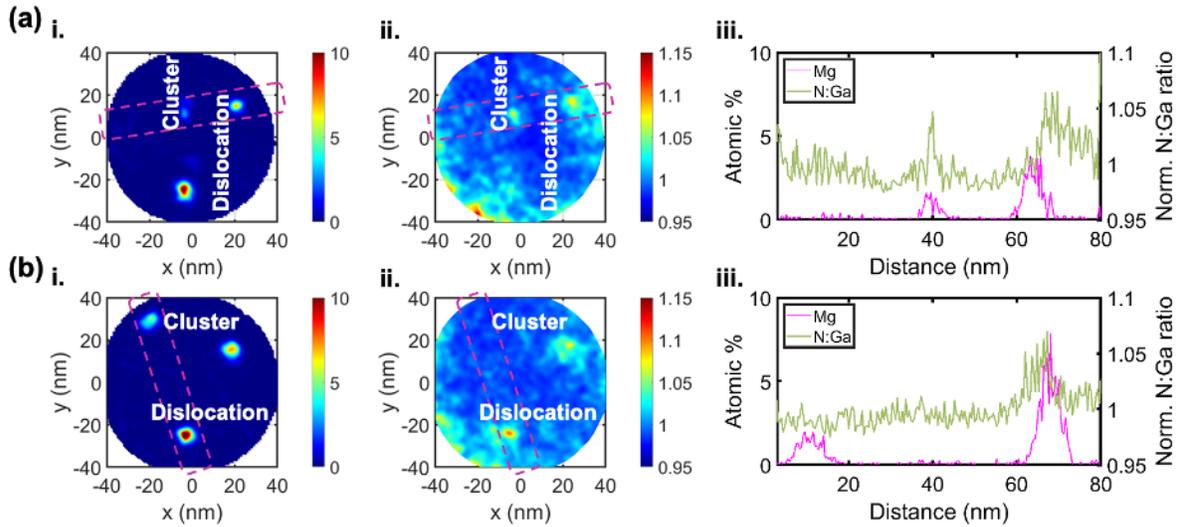

Figure 4

## Supplementary

Figure S1

Figure S1a demonstrates the two regions: (i) with no Mg segregation and (ii) with Mg cluster and dislocation cross-sections. The average value of ΔP was taken every 2 nm along z, between regions (i) and (ii) and is plotted in Figure S1b. The average value was relatively stable along the z direction. Next, ΔP was checked within 2 x 2 x 2 nm voxels within region (i) in Figure S1c and region (ii) in Figure S1d. The average pulse count is homogeneous, aside from a few high (~500) at the very edge of the regions.

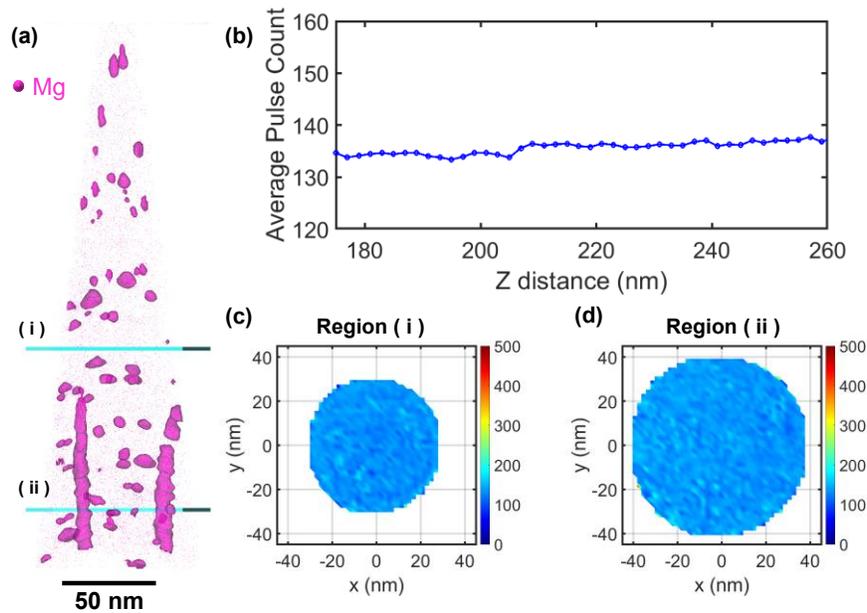

**Figure S1: (a)** 3-D atom map of Mg atoms with boxes indicating 2 nm thick regions: (i) z = 175 and (ii) z = 261. **(b)** average pulse count between regions (i) and (ii). 2D projections of average pulse count taken within 2 x 2 x 2 nm voxels from **(c)** region (i) and **(d)** region (ii).

Figure S2

Figure S2a demonstrates the two regions: (i) with no Mg segregation and (ii) with Mg cluster and dislocation cross-sections. Within the two 2 nm thick regions, the xy-plane was cut into 2 x 2 x 2 cubic nm voxels. Measurements were taken within these voxels for both the density of ions and the relative field evaporation. In Figure S2b, the density of ions from the region without Mg segregation is very similar to Figure S2e, the region containing cluster and dislocation cross-sections. For both, the density of ions shows inhomogeneity; however, the spikes in density do not correlate with any compositional features that we have investigated. The ratio of doubly charged Ga to singly charged Ga was used as a relative measure of the variation in field evaporation, shown in Figure S2c and Figure S2f for the two regions (Bonef, 2017).

Here, both regions show homogeneous distributions. The variation in N:Ga for the region without Mg segregation, shown in Figure S2d, is limited to an edge effect near the lower left quadrant (X: -15, Y: -25). In Figure S2g, there is higher N:Ga near the lower left edge and near both dislocations.

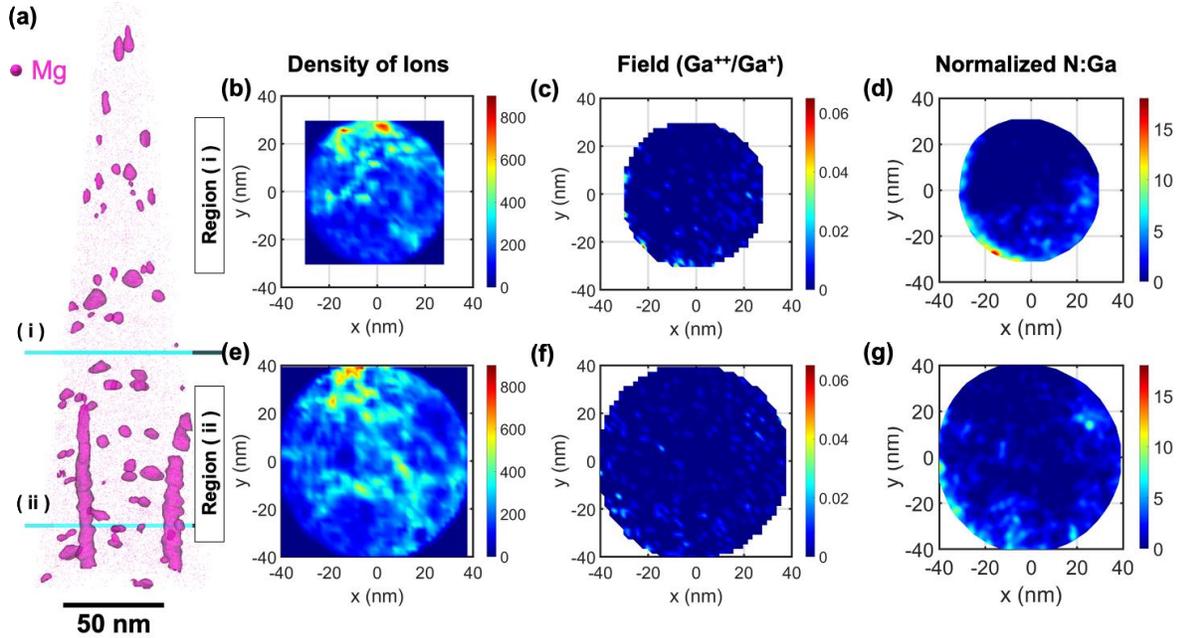

**Figure S2: (a)** 3-D atom map of Mg atoms with boxes indicating 2 nm thick regions: **(i)** z = 175 and **(ii)** z = 261. **(b)** and **(e):** Density of ion (average number of ions per voxel), **(c)** and **(f):** Charge-state-ratio (number of $Ga^{++}$ ions over $Ga^+$ ions per voxel), and **(d)** and **(g):** Normalized N:Ga with mean value equal to 1. The voxel size is 2 x 2 x 2 cubic nanometers.